\documentclass[12pt]{article}
\usepackage{amssymb,amsfonts}
\usepackage{amsmath}
\usepackage{epsf,epsfig}
\begin{document}
\baselineskip 18pt

\title{Low-temperature asymptotics for transverse autocorrelator of the magnetically polarized Ising chain studied by ordinary and nested  Dyson equations}
\author{P.~N.~Bibikov}
\date{\it Russian State Hydrometeorological University, Saint-Petersburg, Russia}

\maketitle

\vskip5mm

\begin{abstract}
Suggesting two versions for the Plakida-Tserkovnikov algorithm breakdown in the low-temperature regime, we derive ordinary
and nested Dyson equations for the transverse autocorrelator of the magnetically polarized Ising chain. Using them
we get the corresponding low-temperature asymptotics for the autocorrelator. We show that the ordinary Dyson
equation results in a correct $o({\rm e}^{-\beta E_{gap}})$ account of the magnon creation process, while the nested Dyson equation additionally
gives the correct $o({\rm e}^{-\beta E_{gap}})$ contribution associated with transitions from magnons to bound two-magnon states.
The obtained result may be useful for the extension of the suggested approach on the polarized $XXZ$-chain.
\end{abstract}

\maketitle

\section{Introduction}

Spectral densities of the form \cite{1}
\begin{equation}
I(\omega,T)=-\frac{1}{\pi(1-{\rm e}^{-\beta\omega})}{\rm Im}\lim_{N\rightarrow\infty}\langle\langle O,O^{\dagger}\rangle\rangle_{\omega},
\end{equation}
are important experimentally measurable dynamical characteristics of magnetic compounds \cite{2,3,4}.
Here $O$ is an appropriate fluctuation operator, $N$ is the number of sites in a magnetic lattice, and for a pair of arbitrary operators ${\cal A}$ and ${\cal B}$ there are two equivalent definitions of the {\it two-real-time retarded Green function} \cite{1}
\begin{equation}
\langle\langle{\cal A},{\cal B}\rangle\rangle_{\omega}\equiv\frac{1}{i}\int_0^{\infty}dt{\rm e}^{izt}\langle[{\cal A}(t),{\cal B}]\rangle=
\frac{1}{i}\int_0^{\infty}dt{\rm e}^{izt}\langle[{\cal A},{\cal B}(-t)]\rangle,\qquad z\equiv\omega+i\epsilon.
\end{equation}
As usual
\begin{equation}
{\cal A}(t)\equiv{\rm e}^{i\hat Ht}{\cal A}{\rm e}^{-i\hat Ht},\qquad\langle{\cal A}\rangle\equiv\frac{1}{Z(T)}{\rm tr}\Big({\rm e}^{-\beta\hat H}{\cal A}\Big),\qquad
Z(T)={\rm tr}\Big({\rm e}^{-\beta\hat H}\Big),
\end{equation}
where $\hat H$ is the corresponding Hamiltonian. From (2) readily follows \cite{1} that
\begin{eqnarray}
z\langle\langle{\cal A},{\cal B}\rangle\rangle_{\omega}=\langle[{\cal A},{\cal B}]\rangle
+\langle\langle{\cal A},[\hat H,{\cal B}]\rangle\rangle_{\omega},\\
z\langle\langle{\cal A},{\cal B}\rangle\rangle_{\omega}=\langle[{\cal A},{\cal B}]\rangle
+\langle\langle[{\cal A},\hat H],{\cal B}\rangle\rangle_{\omega}.
\end{eqnarray}
Besides $\langle\langle{\cal A},{\cal B}\rangle\rangle_{\omega}$ possesses the spectral representation
\begin{equation}
\langle\langle{\cal A},{\cal B}\rangle\rangle_{\omega}=\frac{1}{Z(T,N)}\sum_{\mu,\nu}\frac{{\rm e}^{-\beta E_{\nu}}-{\rm e}^{-\beta E_{\mu}}}{z+E_{\nu}-E_{\mu}}
\langle\nu|{\cal A}|\mu\rangle\langle\mu|{\cal B}|\nu\rangle,
\end{equation}
where $\mu$ and $\nu$ parameterize the eigenbasis of $\hat H$ ($E_{\mu}$ and $E_{\nu}$
are the corresponding energies). If ${\cal A}=O$ and ${\cal B}=O^{\dagger}$, then (6) is reduced to
\begin{equation}
\langle\langle  O,O^{\dagger}\rangle\rangle_{\omega}=\frac{1}{Z(T,N)}\sum_{\mu,\nu}\frac{{\rm e}^{-\beta E_{\nu}}-{\rm e}^{-\beta E_{\mu}}}{z+E_{\nu}-E_{\mu}}
|\langle\nu|O|\mu\rangle|^2.
\end{equation}
According to the well known relation ${\rm Im}(x+i\epsilon)^{-1}=-\pi\delta(x)$, the substitution of (7) into (1) yields
\begin{equation}
I(\omega,T)=\lim_{N\rightarrow\infty}\frac{1}{Z(T,N)}\sum_{\mu,\nu}{\rm e}^{-\beta E_{\nu}}
|\langle\nu|O|\mu\rangle|^2\delta(\omega+E_{\nu}-E_{\mu}).
\end{equation}

The so called $XXZ$ model related to Hamiltonian
\begin{equation}
\hat H^{(XXZ)}=-\sum_{n=1}^N\Big[\frac{J_{\bot}}{2}\Big({\bf S}^+_n{\bf S}^-_{n+1}+{\bf S}^-_n{\bf S}^+_{n+1}\Big)+J_z\Big({\bf S}^z_n{\bf S}^z_{n+1}-\frac{1}{4}\Big)+h\Big({\bf S}_n^z-\frac{1}{2}\Big)\Big],
\end{equation}
is one of the most fundamental models in the theory of 1D magnetic compounds \cite{5}.
Here, ${\bf S}_n^z$ and ${\bf S}_n^{\pm}={\bf S}_n^x\pm i{\bf S}_n^y$
is a triple of spin-1/2 operators attached to the $n$-th site of the chain
\begin{equation}
[{\bf S}_m^z,{\bf S}_n^{\pm}]=\pm\delta_{mn}{\bf S}_n^{\pm},\qquad[{\bf S}_m^+,{\bf S}_n^-]=2\delta_{mn}{\bf S}_n^z,
\end{equation}
and without a loss of generality we postulate
\begin{equation}
h>0.
\end{equation}

Hamiltonian (9) acts in the tensor product of $N$ local ${\mathbb C}^2$ spaces attached to the sites of the chain
${\cal H}={\mathbb C}^2\otimes\dots\otimes{\mathbb C}^2$.
Under the condition (11) supplemented by
\begin{equation}
h+J_z-|J_{\bot}|>0,
\end{equation}
the ground state of (9) has a polarized form
\begin{equation}
|\emptyset\rangle=|\uparrow\rangle\otimes\dots\otimes|\uparrow\rangle,
\end{equation}
where $|\uparrow\rangle$ and $|\downarrow\rangle$ are spin-up and spin-down local polarized states
\begin{equation}
{\bf S}^z|\uparrow\rangle=\frac{1}{2}|\uparrow\rangle,\qquad{\bf S}^z|\downarrow\rangle=-\frac{1}{2}|\downarrow\rangle.
\end{equation}
Moreover under (11) and (12) the spectrum of (9) is gapped and the lowest-energy excitations form the magnon branch \cite{6}
\begin{equation}
|k\rangle={\bf S}^-(-k)|\emptyset\rangle,\qquad\hat H|\emptyset\rangle=E_{magn}(k)|\emptyset\rangle.
\end{equation}
parameterized by crystal momentum $k$. Here
\begin{equation}
{\bf S}(q)\equiv\frac{1}{\sqrt{N}}\sum_{n=1}^N{\rm e}^{-iqn}{\bf S}_n,
\end{equation}
and
\begin{equation}
E_{magn}(k)=h+J_z-J_{\bot}\cos{k},
\end{equation}
so (12) takes the following  form
\begin{equation}
E_{gap}=h+J_z-|J_{\bot}|>0.
\end{equation}.

One of the most important dynamical characteristics of the $XXZ$ chain is its dynamical structure factor (DSF) $S(\omega,q,T)$ which corresponds to
the substitution of $O={\bf S}^+(q)$ and $O^{\dagger}={\bf S}^-(-q)$ into (1). In the the gapped phase (12), (13), the spectral expansion (7) at $T=0$ yields for $S(\omega,q,T)$ a singular delta-function lineshape \cite{7}
\begin{equation}
S(\omega,q,0)=\delta(\omega-E_{magn}(q)).
\end{equation}
This follows from the one-to-one correspondence (17) between the crystal momentum
and the energy of a magnon created from the ground state by the fluctuation operator ${\bf S}^-(-q)$
\begin{equation}
|\emptyset\rangle\xrightarrow{{\bf S}^-(-q)}|q\rangle.
\end{equation}

At $T>0$, due to the existence of thermally activated states, the delta-singularity (19) broadens into a smooth line-shape. This broadening may be well illustrated if one consider, the transition of a thermally
activated magnon with crystal momentum $k_0$ into a scattering (not bound) state of two magnons with crystal momentums $k_1$ and $k_2$
\begin{equation}
|k_0\rangle\xrightarrow{{\bf S}^-(-q)}|k_1,k_2,scatt\rangle.
\end{equation}
Due to the conservation of total crystal momentum, one should have in (21)
\begin{equation}
k_1+k_2=k_0+q.
\end{equation}
According to (22) the delta-function in (8) will give
\begin{equation}
\omega=E_{magn}(k_1)+E_{magn}(k_2)-E_{magn}(k_1+k_2-q).
\end{equation}
But for a given $q$, equation (23) does not fix $\omega$.

From the presented example, follows an instructive idea that the low-temperatures broadening may be well elaborated
by cluster expansion with an account of even one- and two-magnon states. However, as it is well known (and also discussed in \cite{7}),
the delta-peak singularity cannot be removed by direct cluster expansion of (8). An adequate approach for an account of the thermal broadening
in a DSF line-shape based on the Dyson equation for the real two-time Green retarded functions (2) was suggested by N. M. Plakida
in \cite{8,9} and then developed by Yu. A. Tserkovnikov in \cite{10}. Within this approach, a two-real-time retarded Green function is
represented in a continued fraction form. Since in the general case the latter is infinite, one should
effectively break it down in order to get a reduced tractable result. In \cite{7}, this program was realized for evaluation of the low-temperature asymptotics of the transverse magnetic susceptibility
\begin{equation}
\chi(\omega,q,T)=\lim_{N\rightarrow\infty}\langle\langle{\bf S}^+(q),{\bf S}^-(-q)\rangle\rangle_{\omega},
\end{equation}
for the special version of (9) related to $J_z=0$ (the so-called $XX$ model \cite{11}). Following \cite{8,9}, $\chi(\omega,q,T)$ has been represented in the form
\begin{equation}
\chi(\omega,q,T)=\frac{1}{z-E_{magn}(q)-\Sigma(\omega,q,T)},\qquad\Sigma(\omega,q,0)=0,
\end{equation}
associated with the Dyson equation. With the use of one- and two-magnon states the self-energy $\Sigma(\omega,q,T)$ has been evaluated up to the order $o({\rm e}^{-\beta E_{gap}})$.
In the resonance region of $\omega-q$ plane where ${\rm Im}\Sigma(\omega,q,T)>0$,
the line-shape of the DSF corresponding to (25) has a single broad peak.

An extension of this result on the XXZ model (9) seems very desirable. However, in this way, additionally to (20) and
(21), one has to deal with a contribution from a new process, which is a bound state creation
\begin{equation}
|k_0\rangle\xrightarrow{{\bf S}^-(-q)}|k,bound\rangle,
\end{equation}
(spectrum of the $XX$ model is free from bound states \cite{11}). As it will be shown in the present paper, the process (26) is responsible for
a thermally activated peak, which has been discussed over the last fifty years \cite{12,13,14,15,16}.

In order to get an experience in account of (26) before attacking the general Hamiltonian (9), we shall treat here its reduced $(J_{\bot}=0)$ version
\begin{equation}
\hat H^{(Is)}=-\sum_{n=1}^N\Big[J_z\Big({\bf S}^z_n{\bf S}^z_{n+1}-\frac{1}{4}\Big)+h\Big({\bf S}_n^z-\frac{1}{2}\Big)\Big],
\end{equation}
known as the Ising model \cite{17,18}.
The DSF corresponding to (27) is equal to the transverse autocorrelator obtained long ago \cite{17}
\begin{equation}
\langle\langle{\bf S}^+(q),{\bf S}^-(-q)\rangle\rangle_{\omega}=\langle\langle{\bf S}_n^+,{\bf S}_n^-\rangle\rangle_{\omega}.
\end{equation}
Formula (28) immediately follows from the equality $\langle\langle{\bf S}_m^+,{\bf S}_n^-\rangle\rangle_{\omega}=0$ valid for $m\neq n$.
The latter, in its turn, may be readily obtained from the definition (2), (3) under a substitution of (27).
As it has been shown in \cite{17}
\begin{equation}
\langle\langle{\bf S}_n^+,{\bf S}_n^-\rangle\rangle_{\omega}=\frac{A_+(T)}{z-\omega_+}+\frac{A_0(T)}{z-\omega_0}+\frac{A_-(T)}{z-\omega_-},
\end{equation}
where
\begin{equation}
\omega_+=h+J_z,\qquad\omega_0=h,\qquad\omega_-=h-J_z,
\end{equation}
and
\begin{eqnarray}
&&A_{\pm}(T)=\frac{\sigma_1(T)}{2}\pm2\sigma_2(T)+2\sigma_3(T),\qquad A_0(T)=\sigma_1(T)-4\sigma_3(T),\\
&&\sigma_1(T)\equiv\langle{\bf S}^z_n\rangle,\qquad\sigma_2(T)\equiv\langle{\bf S}^z_n{\bf S}^z_{n+1}\rangle,\qquad\sigma_3(T)\equiv\langle{\bf S}^z_{n-1}{\bf S}^z_n{\bf S}^z_{n+1}\rangle,
\end{eqnarray}
(according to translation invariance of (27), all expressions in (32) do not depend on $n$).

Since as in \cite{7} we are interesting in the {\it low-temperature polarized} phase it is instructive to obtain the corresponding asymptotics for the coefficients in (31). This may be done with the use of the low-energy spectrum of (27) and the following representations
\begin{equation}
A_j(T)=\langle F_n^{(j)}\rangle,\qquad j=+,0,-,
\end{equation}
where
\begin{equation}
F^{(\pm)}_n=\frac{1}{2}{\bf S}_n^z\pm{\bf V}_n^z+2{\bf W}_n^z,\qquad F_n^{(0)}={\bf S}_n^z-4{\bf W}_n^z,
\end{equation}
or equivalently
\begin{eqnarray}
&&F^{(+)}_n=2(1-Q_{n-1}){\bf S}^z_n(1-Q_{n+1}),\nonumber\\
&&F^{(0)}_n=2{\bf S}^z_n(Q_{n-1}+Q_{n+1}-2Q_{n-1}Q_{n+1}),\nonumber\\
&&F^{(-)}_n=2Q_{n-1}{\bf S}^z_nQ_{n+1}.
\end{eqnarray}
Here
\begin{equation}
{\bf V}^j_n\equiv({\bf S}_{n-1}^z+{\bf S}_{n+1}^z){\bf S}_n^j,\qquad{\bf W}^j_n\equiv{\bf S}_{n-1}^z{\bf S}_{n+1}^z{\bf S}_n^j,\qquad j=+,-,z,
\end{equation}
and for
\begin{equation}
Q_n\equiv\frac{1}{2}-{\bf S}^z_n,
\end{equation}
one has
\begin{equation}
Q_m|\emptyset\rangle=0,\qquad [Q_m,{\bf S}^-_n]=\delta_{mn}{\bf S}^-_n.
\end{equation}

The spectrum of the Ising model is well known \cite{18} and the actual low-energy states will be the following
\begin{eqnarray}
&&|m\rangle={\bf S}^-_m|\emptyset\rangle,\qquad\hat H|m\rangle=E_1|m\rangle\nonumber\\
&&|m_1,m_2;isol\rangle={\bf S}^-_{m_1}{\bf S}^-_{m_2}|\emptyset\rangle,\qquad m_2-m_1>1,
\nonumber\\
&&\hat H|m_1,m_2;isol\rangle=E_{2,isol}|m_1,m_2;isol\rangle,\nonumber\\
&&|m,m+1;string\rangle={\bf S}^-_m{\bf S}^-_{m+1}|\emptyset\rangle,\nonumber\\
&&\hat H|m,m+1;string\rangle=E_{2,string}|m,m+1;string\rangle\nonumber\\
&&|m,m+1,m+2;string\rangle={\bf S}^-_m{\bf S}^-_{m+1}{\bf S}^-_{m+2}|\emptyset\rangle,\nonumber\\
&&\hat H|m,m+1,m+2;string\rangle=E_{3,string}|m,m+1,m+2;string\rangle.
\end{eqnarray}
Comparing (39) with the corresponding excitations of the $XXZ$ chain \cite{6}, one may conclude that
$|m\rangle$ are the Ising limits of one-magnon states, $|m_1,m_2;isol\rangle$ are the corresponding limits of scattering
two-magnon states. Finally $|m,\dots,m+M;string\rangle$ are related to bound $M$-magnon states.
The corresponding energies
\begin{equation}
E_1=E_{gap}=h+J_z,\qquad E_{M,isol}=M(h+J_z)=ME_1,\qquad
E_{M,string}=Mh+J_z,
\end{equation}
may be readily obtained by elementary calculations. The equality $E_{gap}=E_1$ (which in fact is the reduced form of (18) related to $J_{\bot}=0$)
follows from (40) and (12).

According to (14), (38) and (35), one has
\begin{eqnarray}
&&F^{(+)}_n|\emptyset\rangle=|\emptyset\rangle,\qquad F^{(+)}_n|m\rangle=(1-\delta_{m-1,n}-\delta_{m+1,n}-2\delta_{mn})|m\rangle\nonumber\\ &&F_n^{(0)}|\emptyset\rangle,\qquad F_n^{(0)}|m\rangle=(\delta_{m-1,n}+\delta_{m+1,n})|m\rangle,\nonumber\\ &&F_n^{(-)}|\emptyset\rangle=F_n^{(-)}|m\rangle=0,\qquad F_n^{(-)}|m_1,m_2,isol\rangle=\delta_{m_1,n-1}\delta_{m_2,n+1}|m_1,m_2,isol\rangle.
\end{eqnarray}
So with the use of the notation
\begin{equation}
\zeta\equiv{\rm e}^{-\beta E_{gap}}={\rm e}^{-\beta(h+J_z)},
\end{equation}
(33) and (41) yield
\begin{eqnarray}
&&A_+(T)=\frac{1+(N-4)\zeta+o(\zeta)}{1+N\zeta+o(\zeta)}
=1-4\zeta+o(\zeta),\nonumber\\
&&A_0(T)=\frac{2\zeta+o(\zeta)}{1+o(1)}=2\zeta+o(\zeta),\qquad
A_-(T)=\frac{\zeta^2+o(\zeta^2)}{1+o(1)}=\zeta^2+o(\zeta^2).
\end{eqnarray}
According to (43), the $\omega_+$-peak has a zero activation energy, while the activation energies of $\omega_0$- and $\omega_-$-peaks
are correspondingly $E_{gap}$ and $2E_{gap}$.

A physical interpretation of the expansions (43) becomes elementary under an account of the following relations
\begin{equation}
\omega_+=E_1-0=E_{2,isol}-E_1,\qquad\omega_0=E_{2,string}-E_1\qquad\omega_-=E_{3,string}-E_{2,isol}.
\end{equation}
As it follows from (44), in the few-particle level the $\omega_+$-peak originates from the processes (20) and (21). The
$\omega_0$-peak originates from (26). At last, the $\omega_-$-peak originates from the new process
\begin{equation}
|m,m+2;isol\rangle\xrightarrow{{\bf S}^-(-q)}|m,m+1,m+2;string\rangle.
\end{equation}

In principle, formulas (29)-(32) may be reproduced by the Plakida-Tserkovnikov approach because in this special case,
the corresponding continued fraction is finite. This result, however, is useless for generalization on the
$XXZ$ chain, whose DSF should have an infinite continued fraction representation.
That is why in the present paper, working within the Plakida-Tserkovnikov approach, we try to reproduce the formula (29) and the
low-temperature asymptotics (43) {\it up to the order} $o(\zeta)$ by breaking the continued fraction. This is done by cluster expansion
of the corresponding high-level Green functions (which naturally appear in the Plakida-Tserkovnikov algorithm) according to
the spectral formula (7).

As a result, we show by direct calculation that the correct $o(\zeta)$ approximation for the $A_+(T)$ term in (29)
(which agrees both with (43) and (30)) may be reproduced under the breakdown of the Plakida-Tserkovnikov algorithm after its first cycle.
The corresponding reduction of the continued fraction has the usual form (25) associated with the {\it ordinary} Dyson equation. The $A_0(T)$ term is also reproduced within this approach; however, its denominator does not agree with (29).
The $A_-(T)$ term cannot be obtained within the approach based on the ordinary Dyson equation.
A more complicated reduction is based on the breakdown of the Plakida-Tserkovnikov algorithm after its second cycle.
It yields the nested Dyson equation and gives correct $o(\zeta)$ expansions (which agree both with (43) and (30)) both for
$A_+(T)$ and $A_0(T)$ terms in (29).

\section{Derivation of the ordinary Dyson equation}

\subsection{Extraction of the first order irreducible parts}

Let
\begin{equation}
\theta_0(T)\equiv\frac{\langle[[O_0,\hat H],O_0^{\dagger}]\rangle}{\langle[O_0,O_0^{\dagger}]\rangle},\qquad O_0\equiv O,\qquad
O_0^{\dagger}\equiv O^{\dagger}.
\end{equation}
Since for an arbitrary operator ${\cal A}$ it will be ${\rm Tr}({\rm e}^{-\beta\hat H}[{\cal A},\hat H])=0$, we have
\begin{equation}
\theta_0(T)-\bar\theta_0(T)=\frac{\langle[[O_0,\hat H],O_0^{\dagger}]\rangle-\langle[[O_0^{\dagger},\hat H],O_0]\rangle}{\langle[O_0,O_0^{\dagger}]\rangle}
=\frac{\langle[[O_0,O_0^{\dagger}],\hat H]\rangle}{\langle[O_0,O_0^{\dagger}]\rangle}=0.
\end{equation}
According to (46), a two (temperature-dependent) adjoint operators
\begin{equation}
O_1(T)\equiv[O_0,\hat H]-\theta_0(T)O_0,\qquad O_1^{\dagger}(T)=[\hat H,O_0^{\dagger}]-\theta_0(T)O_0^{\dagger},
\end{equation}
which will be called the {\it first order irreducible parts} of $O_0$ and $O_0^{\dagger}$,
satisfy the {\it first order irreducibility conditions}
\begin{equation}
\langle[O_1(T),O_0^{\dagger}]\rangle=0,\qquad\langle[O_0,O_1^{\dagger}(T)]\rangle=0.
\end{equation}

Substitutions of ${\cal B}=O_0^{\dagger}$ into (4) and ${\cal A}=O_0$ into (5) under an account of (46)-(48) yield
\begin{eqnarray}
&&(z-\theta_0(T))\langle\langle{\cal A},O_0^{\dagger}\rangle\rangle_{\omega}=\langle[{\cal A},O_0^{\dagger}]\rangle
+\langle\langle{\cal A},O_1^{\dagger}(T)\rangle\rangle_{\omega},\\
&&(z-\theta_0(T))\langle\langle O_0,{\cal B}\rangle\rangle_{\omega}=\langle[O_0,{\cal B}]\rangle
+\langle\langle O_1(T),{\cal B}\rangle\rangle_{\omega}.
\end{eqnarray}
Taking now ${\cal A}=O_0$ in (50) and ${\cal B}=O_0^{\dagger}$ in (51), one readily gets
\begin{eqnarray}
(z-\theta_0(T))\langle\langle O_0,O_0^{\dagger}\rangle\rangle_{\omega}=\langle[O_0,O_0^{\dagger}]\rangle
+\langle\langle O_0,O_1^{\dagger}(T)\rangle\rangle_{\omega},\\
(z-\theta_0(T))\langle\langle O_0,O_0^{\dagger}\rangle\rangle_{\omega}=\langle[O_0,O_0^{\dagger}]\rangle
+\langle\langle O_1(T),O_0^{\dagger}\rangle\rangle_{\omega}.
\end{eqnarray}
At the same time, according to (49), the substitution of ${\cal A}=O_1(T)$ into (50) gives
\begin{equation}
(z-\theta_0(T))\langle\langle O_1(T),O_0^{\dagger}\rangle\rangle_{\omega}=
\langle\langle O_1(T),O_1^{\dagger}(T)\rangle\rangle_{\omega}.
\end{equation}

\subsection{Introduction of the first level irreducible Green function}

At this stage, we should express $\langle\langle O_0,O_0^{\dagger}\rangle\rangle_{\omega}$ from $\langle\langle O_1(T),O^{\dagger}_1(T)\rangle\rangle_{\omega}^{(1)}$,
where for two arbitrary operators ${\cal A}$ and ${\cal B}$, the {\it first level irreducible Green function} is defined as
\begin{equation}
\langle\langle{\cal A},{\cal B}\rangle\rangle_{\omega}^{(1)}\equiv\langle\langle{\cal A},{\cal B}\rangle\rangle_{\omega}-
\frac{\langle\langle{\cal A},O_0^{\dagger}\rangle\rangle_{\omega}\langle\langle O_0,{\cal B}\rangle\rangle_{\omega}}{\langle\langle O_0,O_0^{\dagger}\rangle\rangle_{\omega}}.
\end{equation}
From definition (55) immediately follows the {\it first level irreducibility condition}
\begin{equation}
\langle\langle{\cal A}+\xi_0O_0,{\cal B}+\eta_0O_0^{\dagger}\rangle\rangle_{\omega}^{(1)}=\langle\langle{\cal A},{\cal B}\rangle\rangle_{\omega}^{(1)},
\qquad\xi_0,\,\eta_0\in{\mathbb C},
\end{equation}
which, in its turn, gives
\begin{equation}
\langle\langle{\cal A},O_0^{\dagger}\rangle\rangle_{\omega}^{(1)}=\langle\langle O_0,{\cal B}\rangle\rangle_{\omega}^{(1)}=0.
\end{equation}

According to (54)
\begin{equation}
\langle\langle O_1(T),O_1^{\dagger}(T)\rangle\rangle_{\omega}\cdot\langle\langle O_0,O_0^{\dagger}\rangle\rangle_{\omega}=\langle\langle O_1(T),O_0^{\dagger}\rangle\rangle_{\omega}
(z-\theta_0(T))\cdot\langle\langle O_0,O_0^{\dagger}\rangle\rangle_{\omega}.
\end{equation}
The substitution of the right side of (52) into the right side of (58) yields
\begin{equation}
\langle\langle O_1(T),O_1^{\dagger}(T)\rangle\rangle_{\omega}\langle\langle O_0,O_0^{\dagger}\rangle\rangle_{\omega}=\langle\langle O_1(T),O_0^{\dagger}\rangle\rangle_{\omega}
\Big(\langle[O_0,O_0^{\dagger}]\rangle+\langle\langle O_0,O_1^{\dagger}(T)\rangle\rangle_{\omega}\Big),
\end{equation}
or equivalently
\begin{equation}
\langle[O_0,O_0^{\dagger}]\rangle\langle\langle O_1(T),O_0^{\dagger}\rangle\rangle_{\omega}=
\langle\langle O_1(T),O_1^{\dagger}(T)\rangle\rangle_{\omega}^{(1)}\langle\langle O_0,O_0^{\dagger}\rangle\rangle_{\omega}.
\end{equation}

Extracting now $\langle\langle O_1(T),O_0^{\dagger}\rangle\rangle_{\omega}$ in the left side of (60) from (53), one readily gets the ordinary Dyson equation
\begin{equation}
\langle[O_0,O_0^{\dagger}]\rangle\Big((z-\theta_0(T))\langle\langle O_0,O_0^{\dagger}\rangle\rangle_{\omega}-\langle[O_0,O_0^{\dagger}]\rangle\Big)
=\langle\langle O_1(T),O_1^{\dagger}(T)\rangle\rangle_{\omega}^{(1)}\langle\langle O_0,O_0^{\dagger}\rangle\rangle_{\omega},
\end{equation}
or equivalently
\begin{equation}
\langle\langle O,O^{\dagger}\rangle\rangle_{\omega}\equiv\langle\langle O_0,O_0^{\dagger}\rangle\rangle_{\omega}
=\cfrac{\langle[O_0,O_0^{\dagger}]\rangle}{z-\theta_0(T)-\cfrac{\langle\langle O_1(T),O_1^{\dagger}(T)\rangle\rangle_{\omega}^{(1)}}{\langle[O_0,O_0^{\dagger}]\rangle}}\,.
\end{equation}
A correspondence between (62) and the more usual form (25) was studied in \cite{7}.

\section{Derivation of the nested Dyson equation}

\subsection{Equations on the first level irreducible Green functions}

Let ${\cal A}$ and ${\cal B}$ be {\it first order irreducible} operators so that
\begin{equation}
\langle[{\cal A},O_0^{\dagger}]\rangle=\langle[O_0,{\cal B}]\rangle=0.
\end{equation}
The substitution of (63) into (4) and (5) yields
\begin{eqnarray}
&&z\langle\langle O_0,{\cal B}\rangle\rangle_{\omega}=\langle\langle O_0,[\hat H,{\cal B}]\rangle\rangle_{\omega},\\
&&z\langle\langle{\cal A},O_0^{\dagger}\rangle\rangle_{\omega}=\langle\langle[{\cal A},\hat H],O_0^{\dagger}\rangle\rangle_{\omega}.
\end{eqnarray}
Representing now $z\langle\langle{\cal A},{\cal B}\rangle\rangle^{(1)}_{\omega}$ in two equivalent forms
\begin{eqnarray}
&&z\langle\langle{\cal A},{\cal B}\rangle\rangle^{(1)}_{\omega}=
z\langle\langle{\cal A},{\cal B}\rangle\rangle_{\omega}-\frac{z\langle\langle{\cal A},O_0^{\dagger}\rangle\rangle_{\omega}\cdot\langle\langle O_0,{\cal B}\rangle\rangle_{\omega}}{\langle\langle O_0,O_0^{\dagger}\rangle\rangle_{\omega}},
\nonumber\\
&&z\langle\langle{\cal A},{\cal B}\rangle\rangle^{(1)}_{\omega}=
z\langle\langle{\cal A},{\cal B}\rangle\rangle_{\omega}-\frac{\langle\langle{\cal A},O_0^{\dagger}\rangle\rangle_{\omega}\cdot z\langle\langle O_0,{\cal B}\rangle\rangle_{\omega}}{\langle\langle O_0,O_0^{\dagger}\rangle\rangle_{\omega}},
\end{eqnarray}
and utilizing (4), (5), (64) and (65), one readily gets
\begin{eqnarray}
&&z\langle\langle{\cal A},{\cal B}\rangle\rangle^{(1)}_{\omega}=\langle[{\cal A},{\cal B}]\rangle
+\langle\langle{\cal A},[\hat H,{\cal B}]\rangle\rangle_{\omega}
-\frac{\langle\langle{\cal A},O_0^{\dagger}\rangle\rangle_{\omega}\langle\langle O_0,[\hat H,{\cal B}]\rangle\rangle_{\omega}}{\langle\langle O_0,O_0^{\dagger}\rangle\rangle_{\omega}},\nonumber\\
&&z\langle\langle{\cal A},{\cal B}\rangle\rangle^{(1)}_{\omega}=\langle[{\cal A},{\cal B}]\rangle
+\langle\langle[{\cal A},\hat H],{\cal B}\rangle\rangle_{\omega}-\frac{\langle\langle[{\cal A},\hat H],O_0^{\dagger}\rangle\rangle_{\omega}\langle\langle O_0,{\cal B}\rangle\rangle_{\omega}}{\langle\langle O_0,O_0^{\dagger}\rangle\rangle_{\omega}},
\end{eqnarray}
or equivalently
\begin{eqnarray}
&&z\langle\langle{\cal A},{\cal B}\rangle\rangle^{(1)}_{\omega}=\langle[{\cal A},{\cal B}]\rangle+\langle\langle {\cal A},[\hat H,{\cal B}]\rangle\rangle_{\omega}^{(1)},\\
&&z\langle\langle {\cal A},{\cal B}\rangle\rangle^{(1)}_{\omega}=\langle[{\cal A},{\cal B}]\rangle+\langle\langle[{\cal A},\hat H],{\cal B}\rangle\rangle_{\omega}^{(1)}.
\end{eqnarray}
Equations (68) and (69) are the first level analogs of (4) and (5).

\subsection{Extraction of the second order irreducible operator}

After the analogy of (48), we define the {\it second order irreducible part of $O_0$}
\begin{equation}
O_2(T)\equiv[O_1(T),\hat H]-\theta_1(T)O_1(T)-\rho_1(T)O_0,
\end{equation}
where
\begin{equation}
\theta_1(T)=\bar\theta_1(T)=\frac{\langle[[O_1(T),\hat H],O_1^{\dagger}(T)]\rangle}{\langle[O_1(T),O_1^{\dagger}(T)]\rangle},\qquad
\rho_1(T)=\frac{\langle[[O_1(T),\hat H],O_0^{\dagger}]\rangle}{\langle[O_0,O_0^{\dagger}]\rangle}.
\end{equation}
The proof of the relation $\bar\theta_1(T)=\theta_1(T)$ is similar to (47). Let us also prove that
\begin{equation}
\bar\rho_1(T)=\rho_1(T).
\end{equation}
Really, according to (48) and (46),
\begin{equation}
\rho_1(T)=\frac{\langle[[O_1(T),\hat H],O_0^{\dagger}]\rangle}{\langle[O_0,O_0^{\dagger}]\rangle}
=\frac{\langle[[[O_0,\hat H],\hat H],O_0^{\dagger}]\rangle}{\langle[O_0,O_0^{\dagger}]\rangle}-\theta_0^2(T).
\end{equation}
So (72) is equivalent to
\begin{equation}
\langle[O_0,[\hat H,[\hat H,O_0^{\dagger}]]]\rangle=\langle[[[O_0,\hat H],\hat H],O_0^{\dagger}]\rangle.
\end{equation}
The latter, in its turn, follows from (3) and the equality
\begin{eqnarray}
&&[O_0,[\hat H,[\hat H,O_0^{\dagger}]]]-[[[O_0,\hat H],\hat H],O_0^{\dagger}]\nonumber\\
&&=[(O_0O_0^{\dagger}+O_0^{\dagger}O_0),\hat H^2]
-2[(O_0^{\dagger}\hat HO_0+O_0\hat HO_0^{\dagger}),\hat H].
\end{eqnarray}

Equations (70)-(72) result in the following {\it second order irreducibility conditions}
\begin{equation}
\langle[O_2(T),O_1^{\dagger}(T)]\rangle=\langle[O_2(T),O_0^{\dagger}]\rangle=
\langle[O_1(T),O_2^{\dagger}(T)]\rangle=\langle[O_0,O_2^{\dagger}(T)]\rangle=0.
\end{equation}

The following equations
\begin{eqnarray}
&&(z-\theta_1(T))\langle\langle O_1(T),O_1^{\dagger}(T)\rangle\rangle_{\omega}^{(1)}=\langle[O_1(T),O_1^{\dagger}(T)]\rangle
+\langle\langle O_1(T),O_2^{\dagger}(T)\rangle\rangle_{\omega}^{(1)},\quad\\
&&(z-\theta_1(T))\langle\langle O_1(T),O_1^{\dagger}(T)\rangle\rangle_{\omega}^{(1)}=\langle[O_1(T),O_1^{\dagger}(T)]\rangle
+\langle\langle O_2(T),O_1^{\dagger}(T)\rangle\rangle_{\omega}^{(1)},\quad\\
&&(z-\theta_1(T))\langle\langle O_2(T),O_1^{\dagger}(T)\rangle\rangle_{\omega}^{(1)}=
\langle\langle O_2(T),O_2^{\dagger}(T)\rangle\rangle_{\omega}^{(1)},
\end{eqnarray}
may be readily obtained from (68) and (69) in the same manner as (52)-(54) were obtained from (4) and (5) (in former case, however,
one has to additionally account for (57)).

\subsection{Introduction of the second level irreducible Green function}

Let
\begin{equation}
\langle\langle{\cal A},{\cal B}\rangle\rangle_{\omega}^{(2)}\equiv\langle\langle{\cal A},{\cal B}\rangle\rangle_{\omega}^{(1)}-
\frac{\langle\langle{\cal A},O_1^{\dagger}(T)\rangle\rangle_{\omega}^{(1)}\langle\langle O_1(T),{\cal B}\rangle\rangle_{\omega}^{(1)}}{\langle\langle O_1(T),O_1^{\dagger}(T)\rangle\rangle_{\omega}^{(1)}},
\end{equation}
be the {\it second level irreducible Green function}. According to (80) and (56) for two arbitrary ${\cal A}$ and ${\cal B}$ and four arbitrary
complex numbers $\xi_1$, $\xi_0$, $\eta_1$ and $\eta_0$, one has the {\it second level irreducibility property}
\begin{equation}
\langle\langle{\cal A}+\xi_1O_1(T)+\xi_0O_0,{\cal B}+\eta_1O_1^{\dagger}(T)+\eta_0O_0^{\dagger}\rangle\rangle_{\omega}^{(2)}=\langle\langle{\cal A},{\cal B}\rangle\rangle_{\omega}^{(2)}.
\end{equation}
In the same manner as it was done for (62), one may prove that
\begin{equation}
\langle\langle O_1(T),O_1^{\dagger}(T)\rangle\rangle_{\omega}^{(1)}
=\cfrac{\langle[O_1(T),O_1^{\dagger}(T)]\rangle}{z-\theta_1(T)-\cfrac{\langle\langle O_2(T),O_2^{\dagger}(T)\rangle\rangle_{\omega}^{(2)}}{\langle[O_1(T),O_1^{\dagger}(T)]\rangle}}\,.
\end{equation}
By itself, (82) is an ordinary Dyson equation for $\langle\langle O_1(T),O_1^{\dagger}(T)\rangle\rangle_{\omega}^{(1)}$.
At the same time, the combination of (82) and (62) results in the nested Dyson equation for $\langle\langle O,O^{\dagger}\rangle\rangle_{\omega}$
\begin{equation}
\langle\langle O,O^{\dagger}\rangle\rangle_{\omega}
=\cfrac{\langle[O_0,O_0^{\dagger}]\rangle}{z-\theta_0(T)-\cfrac{\langle[O_1(T),O_1^{\dagger}(T)]\rangle}{z-\theta_1(T)
-\cfrac{\langle\langle O_2(T),O_2^{\dagger}(T)\rangle\rangle^{(2)}_{\omega}}{\langle[O_1(T),O_1^{\dagger}(T)]\rangle}}}.
\end{equation}

Although the iteration process may be continued \cite{10}, one has to effectively break it in order to obtain a tractable result. In our further
treatment of the Ising model, we shall suggest two such breakdowns.

\section{Ordinary Dyson equation for the Ising chain}

\subsection{Evaluation of $\theta_0(T)$}

Taking
\begin{equation}
O\equiv O_0={\bf S}^+_n,\qquad O\equiv O_0^{\dagger}={\bf S}^-_n,\qquad\hat H\equiv\hat H^{(Is)},
\end{equation}
and accounting for (31) and (32), one obviously has
\begin{equation}
\langle[O_0,O_0^{\dagger}]\rangle=2\langle{\bf S}^z_n\rangle=A_+(T)+A_0(T)+A_-(T).
\end{equation}
So according to (43),
\begin{equation}
\langle[O_0,O_0^{\dagger}]\rangle=2\langle{\bf S}^z_n\rangle=1-2\zeta+o(\zeta).
\end{equation}
From (84), (27) and (36), it follows that
\begin{equation}
[O_0,\hat H]=h{\bf S}^+_n+J_z{\bf V}^+_n,\qquad [\hat H,O_0^{\dagger}]=h{\bf S}^-_n+J_z{\bf V}^-_n.
\end{equation}
Following (10) and (36)
\begin{equation}
[{\bf V}^+_n,{\bf S}^-_n]=[{\bf S}^+_n,{\bf V}^-_n]=2{\bf V}^z_n.
\end{equation}
From (84), (87), and (88) one readily gets
\begin{equation}
[[O_0,\hat H],O_0^{\dagger}]=[O_0,[\hat H,O^{\dagger}_0]]=2(h{\bf S}^z_n+J_z{\bf V}^z_n),
\end{equation}
and the substitution of (86) and (89) into (46) with account for (30) yields
\begin{equation}
\theta_0(T)=h+J_z\frac{\langle{\bf V}^z_n\rangle}{\langle{\bf S}^z_n\rangle}=\omega_+-J_z\gamma(T),
\end{equation}
where
\begin{equation}
\gamma(T)\equiv\frac{\sigma_1(T)-2\sigma_2(T)}{\sigma_1(T)}
=\frac{A_0(T)+2A_-(T)}{A_+(T)+A_0(T)+A_-(T)}.
\end{equation}
According to (43),
\begin{equation}
\gamma(T)=\frac{2\zeta+o(\zeta)}{1+o(1)}=2\zeta+o(\zeta).
\end{equation}
Hence, the substitution of (92) into (90) gives
\begin{equation}
\theta_0(T)=\omega_+-2J_z\zeta+o(\zeta).
\end{equation}

Substitutions of (87) and (90) into (48) result in
\begin{equation}
O_1(T)=J_z[{\bf V}^+_n+(\gamma(T)-1){\bf S}^+_n],\qquad O_1^{\dagger}(T)=J_z[{\bf V}^-_n+(\gamma(T)-1){\bf S}^-_n].
\end{equation}

\subsection{Breaking down the algorithm}

According to (62), $\langle\langle O_0,O_0^{\dagger}\rangle\rangle_{\omega}$ may be expressed from
$\langle\langle O_1(T),O_1^{\dagger}(T)\rangle\rangle_{\omega}^{(1)}$. At the same time, the definition (55) for
$\langle\langle O_1(T),O_1^{\dagger}(T)\rangle\rangle_{\omega}^{(1)}$ implies the knowledge of $\langle\langle O_0,O_0^{\dagger}\rangle\rangle_{\omega}$.
Due to the gap this vicious circle may be partially broken with the use of special {\it effective} operators
\begin{equation}
O_1^{\rm eff}=J_z({\bf V}^+_n-{\bf S}^+_n),\qquad {O_1^{\rm eff}}^{\dagger}=J_z({\bf V}^-_n-{\bf S}^-_n),
\end{equation}
or equivalently
\begin{equation}
O_1^{\rm eff}=-J_z(Q_{n-1}+Q_{n+1}){\bf S}^+_n,\qquad {O_1^{\rm eff}}^{\dagger}=-J_z(Q_{n-1}+Q_{n+1}){\bf S}^-_n.
\end{equation}

Comparing (94) and (95), one readily gets
\begin{equation}
O_1(T)=O_1^{\rm eff}+J_z\gamma(T)O_0,\qquad O_1^{\dagger}(T)={O_1^{\rm eff}}^{\dagger}+J_z\gamma(T)O_0,
\end{equation}
so, according to (56),
\begin{equation}
\langle\langle O_1(T),O_1^{\dagger}(T)\rangle\rangle_{\omega}^{(1)}=\langle\langle O_1^{eff},{O_1^{\rm eff}}^{\dagger}\rangle\rangle_{\omega}^{(1)}.
\end{equation}
From (96) and (38), immediately follows the {\it first order effectiveness relation}
\begin{equation}
{O_1^{\rm eff}}^{\dagger}|\emptyset\rangle=0\Longleftrightarrow\langle\emptyset|O_1^{\rm eff}=0,
\end{equation}
under which, the spectral expansion (7) yields
\begin{equation}
\langle\langle O_0,{O_1^{\rm eff}}^{\dagger}\rangle\rangle_{\omega}=O(\zeta),\qquad
\langle\langle O_1^{\rm eff},O_0^{\dagger}\rangle\rangle_{\omega}=O(\zeta),\qquad
\langle\langle O_1^{\rm eff},{O_1^{\rm eff}}^{\dagger}\rangle\rangle_{\omega}=O(\zeta).
\end{equation}
At the same time (also according to (7)),
\begin{equation}
\langle\langle O_0,O_0^{\dagger}\rangle\rangle_{\omega}=O(1).
\end{equation}
Under (100) and (101), from (98) and the definition (55), it follows that
\begin{equation}
\langle\langle O_1(T),O_1^{\dagger}(T)\rangle\rangle_{\omega}^{(1)}=
\langle\langle O_1^{\rm eff},{O_1^{\rm eff}}^{\dagger}\rangle\rangle_{\omega}+O(\zeta^2).
\end{equation}
Now, the substitution of (102) into (62) gives the ordinary Dyson equation
\begin{equation}
\langle\langle O,O^{\dagger}\rangle\rangle_{\omega}
=\cfrac{\langle[O_0,O_0^{\dagger}]\rangle}{z-\theta_0(T)-\cfrac{\langle\langle O_1^{\rm eff},{O_1^{\rm eff}}^{\dagger}\rangle\rangle_{\omega}}{\langle[O_0,O_0^{\dagger}]\rangle}+o(\zeta)}\,.
\end{equation}

The right side of (103) may be readily evaluated. Really, according to
(96) and (38),
\begin{eqnarray}
&&O_1^{\rm eff}|m\rangle=O_1^{\rm eff}|m_1,m_2;isol\rangle=0,\nonumber\\
&&O_1^{\rm eff}|m,m+1;string\rangle=-J_z(\delta_{m+1,n}|m\rangle+\delta_{m,n}|m+1\rangle).
\end{eqnarray}
Substituting now (99) and (104) into (7), one gets
\begin{equation}
\langle\langle O_1^{\rm eff},{O_1^{\rm eff}}^{\dagger}\rangle\rangle_{\omega}=\sum_{m,\tilde m}
\frac{\zeta|\langle\tilde m|O_1^{\rm eff}|m,m+1;string\rangle|^2}{z+E_1-E_{2,string}}+o(\zeta).
\end{equation}

From (104), (105), (30) and (40), it follows that
\begin{equation}
\langle\langle O_1^{\rm eff},{O_1^{\rm eff}}^{\dagger}\rangle\rangle_{\omega}=\frac{2J_z^2\zeta}{z-\omega_0}
+o(\zeta).
\end{equation}
So, the substitution of (106) into (103) yields
\begin{equation}
\langle\langle{\bf S}_n^+,{\bf S}_n^-\rangle\rangle_{\omega}
=\cfrac{2\langle{\bf S}^z_n\rangle}{z-\omega_++2J_z\zeta-\cfrac{2J_z^2\zeta}{z-\omega_0}+o(\zeta)}
=\frac{(1-2\zeta+o(\zeta))(z-\omega_0)}{(z-\omega_+)(z-\omega_0+2J_z\zeta)+o(\zeta)},
\end{equation}
or in the polar sum form
\begin{equation}
\langle\langle{\bf S}_n^+,{\bf S}_n^-\rangle\rangle_{\omega}=\frac{1-4\zeta+o(\zeta)}{z-\omega_++o(\zeta)}
+\frac{2\zeta+o(\zeta)}{z-\omega_0+2J_z\zeta+o(\zeta)}.
\end{equation}

We see that the first term in (108), which is responsible for the processes (20) and (21), agrees with the exact formulas (29), (30) and (43).
At the same time, denominator in the second term of (108)
(responsible for the process (26)) is different from the corresponding one in (29). Based on this result, we conclude that the ordinary Dyson equation (103) adequately describes in the $o(\zeta)$ order only the processes related to the creation of a single magnon.

\section{Nested Dyson equation for the Ising chain}

\subsection{Evaluation of $\theta_1(T)$}

According to (71), the evaluation of $\theta_1(T)$ requires the knowledge of $\langle[O_1(T),O_1^{\dagger}(T)]\rangle$
and $\langle[[O_1(T),\hat H],O_1^{\dagger}(T)]\rangle$. In this subsection, we shall successively calculate them up to the order $o(\zeta)$.

\begin{itemize}
\item
According to (94), (49) and (84),
\begin{equation}
\langle[O_1(T),O_1^{\dagger}(T)]\rangle=J_z\langle[{\bf V}^+_n,O_1^{\dagger}(T)]\rangle
=J_z^2(\langle[{\bf V}^+_n,{\bf V}^-_n]\rangle+(\gamma(T)-1)\langle[{\bf V}^+_n,O_0^{\dagger}]\rangle).
\end{equation}
At the same time,
\begin{equation}
\langle[{\bf V}^+_n,O_0^{\dagger}]\rangle=\langle[O_1(T)-(\gamma(T)-1)O_0,O_0^{\dagger}]\rangle=
(1-\gamma(T))\langle[O_0,O_0^{\dagger}]\rangle.
\end{equation}
Hence, (109), (110), (85) and the commutation relation
\begin{equation}
[{\bf V}_n^+,{\bf V}_n^-]={\bf S}_n^z+4{\bf W}_n^z,
\end{equation}
yield
\begin{equation}
\langle[O_1(T),O_1^{\dagger}(T)]\rangle=J_z^2[(1-2(1-\gamma(T))^2)\sigma_1(T)+4\sigma_3(T)],
\end{equation}
or, according to (91) and (31),
\begin{eqnarray}
&&\langle[O_1(T),O_1^{\dagger}(T)]\rangle=J_z^2[\sigma_1(T)-4(1-\gamma(T))\sigma_2(T)+4\sigma_3(T)]\nonumber\\
&&=2J_z^2[2\gamma(T)\sigma_2(T)+A_-(T)].
\end{eqnarray}
Taking now into account that according to (32) $\sigma_2(0)=1/4$ and using expansions (43) and (92), one readily gets from (113)
\begin{equation}
\langle[O_1(T),O_1^{\dagger}(T)]\rangle=2J_z^2\zeta+o(\zeta).
\end{equation}

\item
According to (84), (87) and (94),
\begin{equation}
[{\bf S}_n^+,\hat H]=O_1(T)+[h+(1-\gamma(T))J_z]O_0.
\end{equation}
In the same manner, the formula
\begin{equation}
[{\bf V}^+_n,\hat H]=h{\bf V}^+_n+J_z\Big(\frac{1}{2}{\bf S}^+_n+2{\bf W}^+_n\Big),
\end{equation}
(which directly follows from (10), (27) and (36)) may be represented in the following form
\begin{equation}
[{\bf V}^+_n,\hat H]=2J_z{\bf W}^+_n+\frac{h}{J_z}O_1(T)+\Big((1-\gamma(T))h+\frac{J_z}{2}\Big)O_0.
\end{equation}
So, according to (94) and (49),
\begin{equation}
\langle[[O_1(T),\hat H],O_1^{\dagger}(T)]\rangle=(h+(\gamma(T)-1)J_z)\langle[O_1(T),O_1^{\dagger}(T)]\rangle
+2J_z^2\langle[{\bf W}_n^+,O_1^{\dagger}(T)]\rangle,
\end{equation}
and the substitution of (118) into (71) results in
\begin{equation}
\theta_1(T)=h+(\gamma(T)-1)J_z+2J_z^2\frac{\langle[{\bf W}_n^+,O_1^{\dagger}(T)]\rangle}{\langle[O_1(T),O_1^{\dagger}(T)]\rangle}.
\end{equation}

\item

Using definitions (94) and (36), one may readily check that
\begin{equation}
2[{\bf W}_n^+,O_1^{\dagger}(T)]=J_z({\bf V}^z_n-4(1-\gamma(T)){\bf W}^z_n).
\end{equation}
Hence
\begin{equation}
2\langle[{\bf W}_n^+,O_1^{\dagger}(T)]\rangle=2J_z[\sigma_2(T)-2(1-\gamma(T))\sigma_3(T)].
\end{equation}
Taking now into account that according to (31) and (91)
\begin{eqnarray}
&&4(1-\gamma(T))\sigma_3(T)=(1-\gamma(T))[2A_-(T)-\sigma_1(T)+4\sigma_2(T)]\nonumber\\
&&=2[(1-\gamma(T))A_-(T)+(1-2\gamma(T))\sigma_2(T)],
\end{eqnarray}
one readily gets from (121)
\begin{equation}
2\langle[{\bf W}_n^+,O_1^{\dagger}(T)]\rangle=2J_z[2\gamma(T)\sigma_2(T)+(\gamma(T)-1)A_-(T)].
\end{equation}
Now, the comparison between (113) and (123) yields
\begin{equation}
2J_z\langle[{\bf W}_n^+,O_1^{\dagger}(T)]\rangle=\langle[O_1(T),O_1^{\dagger}(T)]\rangle+2J_z^2(\gamma(T)-2)A_-(T).
\end{equation}
Finally, the substitution of (124) into (119) results in
\begin{equation}
\theta_1(T)=\omega_0+J_z\gamma(T)+2J_z^3\frac{(\gamma(T)-2)A_-(T)}{\langle[O_1(T),O_1^{\dagger}(T)]\rangle},
\end{equation}
and with the use of expansions (92), (43) and (114), one has
\begin{equation}
\theta_1(T)=\omega_0+o(\zeta).
\end{equation}

\end{itemize}

\subsection{Breaking down the algorithm}

According to (84) and (94), operators ${\bf S}_n^+$ and ${\bf V}_n^+$ are linear combinations of $O_0$ and $O_1(T)$.
So, using (94), (115), (117
) and (81), one may conclude that for two arbitrary
complex numbers $\xi_1$ and $\xi_0$, one has
\begin{equation}
\langle\langle O_2(T),O_2^{\dagger}(T)\rangle\rangle^{(2)}_{\omega}=\langle\langle\tilde O_2(T),\tilde O_2^{\dagger}(T)\rangle\rangle^{(2)}_{\omega}.
\end{equation}
where
\begin{equation}
\tilde O_2(T)=2J_z^2{\bf W}_n^++\xi_1{\bf V}_n^++\xi_0{\bf S}_n^+.
\end{equation}
As a result,
\begin{equation}
\langle\langle O_2(T),O_2^{\dagger}(T)\rangle\rangle^{(2)}_{\omega}=
\langle\langle O_2^{\rm eff},{O_2^{\rm eff}}^{\dagger}\rangle\rangle^{(2)}_{\omega},
\end{equation}
where
\begin{equation}
O_2^{\rm eff}=J_z^2\Big(2{\bf W}_n^++\frac{1}{2}{\bf S}_n^+-{\bf V}_n^+\Big),
\end{equation}
or, according to (37),
\begin{equation}
O_2^{\rm eff}=2J_z^2Q_{n-1}{\bf S}_n^+Q_{n+1}.
\end{equation}
From (131) and (38), immediately follows the {\it second order effectiveness relation}
\begin{equation}
{O_2^{\rm eff}}^{\dagger}|\emptyset\rangle={O_2^{\rm eff}}^{\dagger}|m\rangle=0,
\end{equation}
and, moreover,
\begin{equation}
{O_2^{\rm eff}}^{\dagger}|m,m+1\rangle={O_2^{\rm eff}}^{\dagger}|m_1,m_2;isol\rangle=0,\qquad m_2-m_1>2.
\end{equation}
Under (132) spectral expansion (7) yields
\begin{equation}
\langle\langle O_j^{\rm eff},{O_2^{\rm eff}}^{\dagger}\rangle\rangle_{\omega}=O(\zeta^2),\qquad
\langle\langle O_2^{\rm eff},{O_j^{\rm eff}}^{\dagger}\rangle\rangle_{\omega}=O(\zeta^2),\qquad
j=0,1,2,
\end{equation}
(where $O_0^{\rm eff}\equiv O_0$).
In its turn, at $j=0$ the substitution of (134) into (55) with account for (101) gives
\begin{equation}
\langle\langle O_2^{\rm eff},{O_2^{\rm eff}}^{\dagger}\rangle\rangle^{(1)}_{\omega}=\langle\langle O_2^{\rm eff},{O_2^{\rm eff}}^{\dagger}\rangle\rangle_{\omega}
+O(\zeta^4)=O(\zeta^2).
\end{equation}
In similar manner, from (100), (101) and (134), it follows that
\begin{eqnarray}
\langle\langle O_2^{\rm eff},{O_1^{\rm eff}}^{\dagger}\rangle\rangle^{(1)}_{\omega}
=\langle\langle O_2^{\rm eff},{O_1^{\rm eff}}^{\dagger}\rangle\rangle_{\omega}+O(\zeta^3)=O(\zeta^2),\nonumber\\
\langle\langle O_1^{\rm eff},{O_2^{\rm eff}}^{\dagger}\rangle\rangle^{(1)}_{\omega}
=\langle\langle O_1^{\rm eff},{O_2^{\rm eff}}^{\dagger}\rangle\rangle_{\omega}+O(\zeta^3)=O(\zeta^2).
\end{eqnarray}

According to (106) and (136),

\begin{equation}
\frac{\langle\langle O_2^{\rm eff},{O_1^{\rm eff}}^{\dagger}\rangle\rangle^{(1)}_{\omega}\langle\langle O_1^{eff},{O_2^{\rm eff}}^{\dagger}\rangle\rangle^{(1)}_{\omega}}
{\langle\langle O_1^{\rm eff},{O_1^{\rm eff}}^{\dagger}\rangle\rangle^{(1)}_{\omega}}=O(\zeta^3).
\end{equation}
Finally, the substitution of (135) and (137) into (80) with account for (129) yields
\begin{equation}
\langle\langle O_2(T),O_2^{\dagger}(T)\rangle\rangle^{(2)}_{\omega}=\langle\langle O_2^{\rm eff},{O_2^{\rm eff}}^{\dagger}\rangle\rangle_{\omega}
+o(\zeta^2).
\end{equation}

According to (133), the evaluation of $\langle\langle O_2^{\rm eff},{O_2^{\rm eff}}^{\dagger}\rangle\rangle_{\omega}$ in the $o(\zeta^2)$ order needs
only to take into account the process (45). From (39) and (131) one has
\begin{equation}
{O_2^{\rm eff}}^{\dagger}|m_1,m_2,isol\rangle=2J_z^2\delta_{m_1,n-1}\delta_{m_2,n+1}|n-1,n,n+1\rangle.
\end{equation}
A substitution of (139) into (7) gives
\begin{equation}
\langle\langle O_2^{\rm eff},{O_2^{\rm eff}}^{\dagger}\rangle\rangle_{\omega}=\frac{4J_z^4\zeta^2}{z+E_{2,isol}-E_{3,string}}
+o(\zeta^2).
\end{equation}
So, according to (40) and (138),
\begin{equation}
\langle\langle O_2(T),O_2^{\dagger}(T)\rangle\rangle_{\omega}^{(2)}=\frac{4J_z^4\zeta^2}{z-\omega_-}
+o(\zeta^2).
\end{equation}
Finally, the substitution of (86), (93), (126) and (141) into (83) results in
\begin{equation}
\langle\langle{\bf S}_n^+,{\bf S}_n^-\rangle\rangle_{\omega}
=\cfrac{1-2\zeta+o(\zeta)}{z-\omega_++2J_z\zeta+o(\zeta)-\cfrac{2J_z^2\zeta+o(\zeta)}{z-\omega_0
-\cfrac{2J_z^2\zeta+o(\zeta)}{z-\omega_-+o(\zeta)}}}.
\end{equation}
After a rather cumbersome calculation, (142) may be transformed into a polar sum form
\begin{equation}
\langle\langle{\bf S}_n^+,{\bf S}_n^-\rangle\rangle_{\omega}=\frac{1-4\zeta+o(\zeta)}{z-\omega_++o(\zeta)}
+\frac{2\zeta+o(\zeta)}{z-\omega_0+o(\zeta)}-\frac{4\zeta^3+o(\zeta^3)}{z-\omega_--2J\zeta+o(\zeta)},
\end{equation}
which, according to (43), reproduces the first two terms of (29) up to the order $o(\zeta)$.

\section{Summary and discussion}

In the present paper, continuing the line of research which has been suggested in \cite{7}, we studied the transverse autocorrelator of the magnetically polarized Ising chain within the Plakida-Tserkovnikov algorithm \cite{8,9,10} reduced by two versions of its breakdown. The
first one yields the ordinary Dyson equation (62) \cite{7}, while the second version - the nested one (83). The obtained results were compared with the exact formula (29) \cite{17}. It has been shown that in the $o({\rm e}^{-\beta E_{gap}})$ order, the obtained ordinary Dyson equation completely reproduces only one
term of the exact autocorrelator. The latter is responsible for the creation of a single magnon. At the same time, the nested Dyson equation additionally reproduces in the $o({\rm e}^{-\beta E_{gap}})$ order the term responsible for the transformation of a single magnon into a pair of neighboring magnons
(an analog of a bound two-magnon state). The third term in the exact formula (29) \cite{17}, which is related to the transition from a pair of isolated magnons (analog of a scattering two-magnon state) into a triple of neighboring magnons (an analog of a bound three-magnon state) and has the activation energy ${\rm e}^{-2\beta E_{gap}}$, cannot be correctly reproduced within the approaches based on (62) or (83).

The main aim of the paper was to suggest the convenient breakdowns and to compare the obtained results with those of the exact formula (29) \cite{17}.
That is why some details of the obtained results are quite surprising and not totally clear for the author.
\begin{itemize}
\item Despite the fact that all the calculations in Sect. 4 were performed in the $o({\rm e}^{-\beta E_{gap}})$ order, the obtained second term in (108) does not agree (within this order) with the corresponding term in the exact formula (29).

\item Formula (108), which correctly describes in the $o({\rm e}^{-\beta E_{gap}})$ order only the (20) and (21) processes, was obtained from
(103). The latter however contains in the denominator the term $\frac{\langle\langle O_1^{\rm eff},{O_1^{\rm eff}}^{\dagger}\rangle\rangle_{\omega}}{\langle[O_0,O_0^{\dagger}]\rangle}$, which accounts for the process (26). But if we remove this term from
(103) then the resulting formula analogous to (108) will be incorrect even for the processes (20) and (21).

\item For the evaluation of the second term in (143) originated from transitions (26) of a single magnon into a string of two neighboring magnons, one has to account for the
process (45) related to the transition of two isolated (scattering) magnons into a string of three neighboring magnons (three-magnon bound state). But contrary to the former one, which has the activation energy $E_{gap}$, the latter one has the energy $2E_{gap}$.
\end{itemize}

We suppose that the calculations developed in this paper will be a guideline for the future investigation of the $XXZ$-chain.

The author is grateful to S. B. Rutkevich for a helpful discussion.


\begin{thebibliography}{18}
\bibitem{1} Rudou Yu G 2011 The Bogoliubov-Tyablikov Green's function method in the quantum theory of magnetism, Theoret. and Math. Phys. {\bf168} 1318-1329
\bibitem{2} Ajiro J 2003 ESR experimens on quantum spin systems, J. Phys. Soc. Jpn., Suppl. {\bf B72}, 12
\bibitem{3} P. Lemmens, G. $\rm G\ddot untherodt$, C. Gros, Magnetic light scattering in low-dimensional quantum spin systems, Phys. Rep. {\bf375}, 1-103 (2003)
\bibitem{4} Zaliznyak I A, Lee S H 2005 Magnetic Neutron Scatttering, in {\it Modern Techniques for Characterizing Magnetic Materials}, Ed. Zhu Y (Springer, Heidelberg)
\bibitem{5} Mikeska H J, Kolezhuk A K 2004 One-dimensional magnetism Lect. Notes in Phys. {\bf645} 1-83
\bibitem{6} Bibikov P N 2015 Second cluster integral from the spectrum of an infinite $XXZ$ chain, Annals of Phys. {\bf354}, 705-714
\bibitem{7} Bibikov P N 2020 Low-temperature asymptotic of the transverse dynamical structure factor for a magnetically
        polarized $XX$ chain Journ. of Stat. Mech. Theor. and Exper. 073106
\bibitem{8} Plakida N M 1973 Dyson equation for Heisenberg ferromagnet, Phys. Lett. A {\bf 43} 481-482
\bibitem{9} Plakida N M 2011 The two-time Green's function and the diagram technique, Theoret. and Math. Phys. {\bf168} 1303-1317
\bibitem{10} Tserkovnikov Yu A 1981 Method of solving infinite systems of equations for two-time temperature Green functions, Theoret. and Math. Phys. {\bf49} 993-1002
\bibitem{11} Colomo F, Izergin A G, Korepin V E, Tognetti V 1993 Temperature correlation functions in the XX0 Heisenberg chain Theoret. Math. Phys. {\bf94} 11-38
\bibitem{12} Villain J 1975 Propagative spin relaxation in the Ising-like antiferromagnetic linear chain, Physica{\bf79B} 1-12
\bibitem{13} James A J A, Goetze W D, Essler F H D 2009 Finite temperature dynamical structure factor of the Heisenberg-Ising chain, Phys. Rev. B {\bf79} 214408
\bibitem{14} Goetze W D, Karahasanovic, Essler F H L 2010 Low-temperature dynamical structure factor of the spin-$\frac{1}{2}$
    Heisenberg ladder, Phys. Rev. {\bf B82} 104417
\bibitem{15} Brockmann M, $\rm G\ddot ohmann$ F, Karbach M, ${\rm Kl\ddot umper}$ A, $\rm Wei\ss e$ 2012 On the absorption of microwaves by the one-dimensional spin-1/2 Heisenberg-Ising magnet, Phys. Rev. {\bf B85} 134438
\bibitem{16} A. K. Bera, B. Lake, F. H. D. Essler, L. Vanderstraeten, C. Hubig, U. $\rm Schollw\ddot ock$, A. T. M. N. Islam, A. Schneidewind, D. L. Quintero-Castro 2017 Spinon confinement in a quasi one dimensional anisotropic Heisenberg magnet, Phys. Rev. {\bf B96}, 054423
\bibitem{17} Oguchi T, Ono I 1966 Theory of the Ising ferromagnet using the Green function method, Progr. Theor. Phys. {\bf35} 998-1009
\bibitem{18} Baxter R J 1982 Exactly Solved Models in Statistical Mechanics, Academic Press, London
\end{thebibliography}
\end{document}